\documentclass[usenatbib,onecolumn,useAMS]{mn2e}  
\usepackage{times}
\usepackage{epsf}

\newcommand{\di}{\rho_{i}}
\newcommand{\df}{\rho_{f}}
\newcommand{\dip}{\rho_{i}'}
\newcommand{\sigi}{\sigma_{i}}
\newcommand{\sigf}{\sigma_{f}}
\newcommand{\ri}{r_{i}}
\newcommand{\rf}{r_{f}}
\newcommand{\zi}{z_{i}}
\newcommand{\zf}{z_{f}}

\title[Dynamical response to supernova-induced gas removal]{
Dynamical response to supernova-induced gas removal in two-component spherical galaxies}

\author[M. Nagashima and Y. Yoshii]{Masahiro Nagashima$^{1}$
\thanks{E-mail: masa@th.nao.ac.jp} and Yuzuru Yoshii$^{2,3}$\\ 
$^{1}$National Astronomical Observatory, Mitaka, Tokyo 181-8588, Japan\\ 
$^{2}$Institute of Astronomy, School of Science, The University of
Tokyo, Mitaka, Tokyo 181-0015, Japan\\
$^{3}$Research Center for the Early Universe, School of Science, The
University of Tokyo, Bunkyo-ku, Tokyo 113-0033, Japan}

\begin{document}   
\maketitle   

\begin{abstract}   
 We investigate dynamical response on size and velocity dispersion to
 mass loss by supernovae in formation of two-component spherical
 galaxies composed of baryon and dark matter.  Three-dimensional
 deprojected de Vaucouleurs-like and exponential-like profiles for
 baryon, embedded in truncated singular isothermal and homogeneous
 profiles for dark matter, are considered.  As a more realistic case, we
 also consider a dark matter profile proposed by Navarro, Frenk \&
 White.  For simplicity we assume that dark matter distribution is not
 affected by mass loss and that the change of baryonic matter
 distribution is homologous.  We found that the degree of the response
 depends on the fraction of dark matter in the region where baryon is
 distributed, so that dwarf spheroidal galaxies would be affected even
 in a dark halo if they are formed by galaxy mergers in the envelope of
 the dark halo.  Our results suggest that this scenario, combined with
 dynamical response, would make not only the observed trends but the
 dispersed characteristics of dwarf spheroidals.
\end{abstract}   
   

\section{INTRODUCTION}   
Elliptical galaxies have provided important information on the formation
and evolution of galaxies.  So far some scaling relations among
photometric properties and global structural parameters have been
observed, such as the colour--magnitude relation \citep[e.g.,][]{b59},
the velocity dispersion--magnitude relation \citep{fj76}, and the
surface brightness--size relation \citep{k77, kow83}.  These relations
are regarded as clues to our understanding of galaxy formation.

It is widely believed that heating up and sweeping out the galactic gas
by multiple supernova (SN) explosions play a key role on the formation
of dwarf elliptical/spheroidal galaxies.  In a traditional framework of
monolithic cloud collapse scenario for formation of ellipticals, such
multiple SNe cause a galactic wind to halt subsequent star formation and
chemical enrichment \citep{l69, i77, s79, ay86, ay87, ka97}.  Since a
large amount of gas is expelled by the galactic wind, the
self-gravitating system expands after blowing up the wind
\citep[e.g.,][]{h80, m83, v86}.  \citet{ya87} found that many observed
properties of ellipticals can be reproduced by the galactic wind model
with the evolutionary population synthesis technique.  They noticed that 
theoretical metallicity of galaxies, to be compared with observation, 
should be an average of metallicities of constituent stars weighted by 
their luminosities.  On the other hand, \citet{ds86} considered a model 
in which galaxies are embedded in dominant dark haloes.  They claimed 
that structural properties of luminous matter would not be affected by 
the gas removal if dark matter dominates the gravitational potential.  
Their model can also reproduce the observed relations, although they did 
not take into account the evolutionary population synthesis.  Recently, 
\citet{z02} argued the dynamical limit on occurrence of galactic wind 
given by the existence of sufficiently dense dark halo and the absence of 
intergalactic/escaping globular clusters, and concluded that only about 
4 per cent of the total mass of the galaxy should be lost by the wind.

In contrast to the monolithic cloud collapse scenario, recent studies on
the cosmological structure formation suggest the hierarchical clustering
scenario based on a cold dark matter (CDM) model.  In this scenario, the
CDM dominates the Universe gravitationally and objects or dark haloes
continuously cluster and merge together, so that larger objects form via
mergers of smaller objects.  Thus, simple galactic wind models must be
modified in order to be consistent with the cosmological structure
formation.  Based on the hierarchical clustering scenario, semi-analytic
models (SAMs) of galaxy formation have been developed
\citep[e.g.,][]{kwg93, c94, c00, sp99, ngs99, n01, n02}.  These models
well reproduce many statistical observables of galaxies such as
luminosity function, colour distribution, and gas fraction.
\citet{kc98} and \citet{ng01} have shown that their SAMs taking into
account metal enrichment can reproduce the observed colour--magnitude
relation of elliptical galaxies in clusters of galaxies and that this
relation reflects the mass-metallicity relation, as pointed out by using
the galactic wind model \citep{ka97}.  Thus, as a next step, it is
valuable to investigate not only the photometric properties but
structural properties such as size and velocity dispersion.  Moreover,
\citet{on01a, on01b} found that radial gradients of fractional number,
colour, and star formation rate of ellipticals in clusters are also
reproduced by using the SAM associated with an $N$-body simulation.

Because a large amount of mass loss might occur during the formation of
ellipticals, in both scenarios of monolithic cloud collapse and
hierarchical clustering, it is important to construct a formalism on the
dynamical response in two-component galaxies consisting of baryon and
dark matter.  While self-consistent equilibrium models of such
two-component elliptical galaxies have been analysed by \citet{ys87} 
and \citet{cp92}, we focus on the dynamical response of baryon
within a dark matter halo.  In this paper we consider simple models of
two-component galaxies in which the baryon having either de
Vaucouleurs or exponential profile is embedded in the dark matter having
either isothermal or homogeneous profile.  Moreover, we also consider the
so-called NFW profile of dark matter that is recently suggested by
$N$-body simulations \citep{nfw}.  

This paper is outlined as follows.  In section 2, general formulation
for the dynamical response on size and velocity dispersion to mass loss
is presented.  In section 3 the dynamical response for particular choices 
of density distributions mentioned above is investigated.  Section 4
summarizes the results of this paper.  The details of models and
derivations are given in Appendices.

\section{EQUATIONS OF DYNAMICAL RESPONSE IN TWO-COMPONENT GALAXIES}
We assume that the velocity fields of gas, stars and dark matter are
non-rotating, homogeneous and isotropic characterized by a single
parameter of velocity dispersion $\sigma$, and that the density 
distribution of baryon changes homologously.  For simplicity, it is 
also assumed that the density distribution of dark matter does not 
change during the mass loss.  This assumption will be justified for 
global structural parameters of dark matter because the total mass of 
dark matter dominates over baryon by a factor of 10 or so.  Although 
the following results will be slightly modified in a dense region of
baryonic matter because of non-negligible dynamical response on the
local dark matter distribution, the resulting change of baryonic matter
should be intermediate between the baryon-dominated and dark
matter-dominated cases.  We will study the dynamical response in such
limiting cases in section 2.2.

The response depends on the timescale of gas removal between two extreme
cases of adiabatic (slow) and instantaneous (rapid) gas removal.  When
the timescale is longer than the dynamical timescale of the system, it
changes adiabatically in quasi-equilibrium.  In this case the total
change is a sum of consecutive infinitesimal changes.  On the other
hand, when the timescale of gas removal is much shorter than the
dynamical timescale, the system drastically changes and it is possible
that our assumption of homologous change might be broken.  Under
realistic circumstances, as hydrodynamical simulations suggest, these
timescales are almost similar \citep[e.g.,][]{mytn, myn, mfm}.
Therefore, keeping in mind that the adiabatic gas removal is more
consistent with the situation under our assumptions, we consider both
adiabatic and instantaneous gas removal below.

\subsection{Formulation}
A kinetic energy $T$ due to velocity dispersion and a gravitational
potential energy $W$ of baryonic component in a dark matter halo are
written as
\begin{eqnarray}
T&=&\frac{1}{2}p\rho_{b} r_{b}^{3}\sigma^{2},\\
W&=&-a\rho_{b}^{2}r_{b}^{5}-b\rho_{b}\rho_{d} r_{b}^{3}r_{d}^{2}f(z),
\label{eqn:genpot}
\end{eqnarray}
respectively, where $p, a$ and $b$ are constants dependent on density
distribution, $\rho_{b}$ and $\rho_{d}$ are characteristic densities of
baryon and dark matter, respectively, $\sigma$ is velocity dispersion of
baryon, $r_{b}$ and $r_{d}$ are characteristic radii of baryon and dark
matter, respectively, $z$ is a radius $r_{b}$ normalized by $r_{d}$,
$z\equiv r_{b}/r_{d}$, and $f(z)$ is a function dependent on density
distribution.  Note that variables related to dark matter, with 
subscript $d$, do not change during the mass loss.  From the virial
theorem, we obtain
\begin{equation}
p\sigi^{2}=a\di\ri^{2}+b\rho_{d}r_{d}^{2}f(\zi),
\label{eqn:viriali}
\end{equation}
\begin{equation}
p\sigf^{2}=a\df\rf^{2}+b\rho_{d}r_{d}^{2}f(\zf),
\label{eqn:virialf}
\end{equation}
where subscripts $i$ and $f$ stand for the initial and final states,
that is, before and after the mass loss, respectively, for the baryonic
component.  Just after the mass loss, only the baryon density changes
from $\di$ to $\di'$ without any changes of other variables.  Then,
conserving its energy and mass, the system virializes again quickly to
the final state,
\begin{equation}
\frac{1}{2}p\dip\ri^{3}\sigi^{2}-a\dip^{2}\ri^{5}-b\dip\rho_{d}\ri^{3}r_{d}^
{2}f(\zi)=
\frac{1}{2}p\df\rf^{3}\sigf^{2}-a\df^{2}\rf^{5}-b\df\rho_{d}\rf^{3}r_{d}^{2}
f(\zf).
\end{equation}
After the mass loss the total mass of baryon also conserves, then 
$\dip\ri^{3}=\df\rf^{3}$.  By using this relation, the above equation is 
transformed to
\begin{equation}
\frac{p}{2}(\sigf^{2}-\sigi^{2})-a\df\rf^{2}\left(1-\frac{\rf}{\ri}\right)-b
\rho_{d}r_{d}^{2}[f(\zf)-f(\zi)]=0.
\end{equation}
Using the virial relation, $\sigi$ and $\sigf$ can be eliminated as
\begin{equation}
1+\frac{\df}{\di}\left[\left(\frac{\zf}{\zi}\right)^{2}-2\left(\frac{\zf}{\zi}\right)^{3}\right]
+\frac{b\rho_{d}}{a\di\zi^{2}}[f(\zf)-f(\zi)]=0.
\label{eqn:diff2}
\end{equation}
When the gas removal occurs instantaneously, the density at the final
state is given by the above equation,
\begin{equation}
 \frac{y_{f}}{y_{i}}=\frac{
1+b[f(z_{f})-f(z_{i})]/ay_{i}z_{i}^{2}}
{(z_{f}/z_{i})^{2}(2z_{f}/z_{i}-1)},
\label{eqn:inst}
\end{equation}
where $y_{i,f}=\rho_{i,f}/\rho_{d}$, respectively.  On the other hand,
when the gas removal is sufficiently slow, the total change is a sum
of consecutive infinitesimal changes, for which $y_{f}=y_{i}+dy$ and
$z_{f}=z_{i}+dz$.  Thus, linearizing the above equation, we obtain
\begin{eqnarray}
\frac{dy}{dz}&=&-4\frac{y}{z}+\frac{b}{az^{2}}\frac{df(z)}{dz},
\label{eqn:diff}
\end{eqnarray}
where subscript $i$ is omitted for simplicity.  This corresponds to
the case of adiabatic change.  The first term in the above equation is
obtained for the self-gravitational potential of baryon.  The second term
gives the modification for baryon embedded in dark matter.  Solving this 
differential equation, we obtain
\begin{equation}
 y=\frac{C}{z^{4}}+q(z),
\end{equation}
where
\begin{equation}
q(z)\equiv\frac{b}{az^{4}}\int_{0}^{z}t^{2}\frac{df(t)}{dt}dt,
\label{eqn:q}
\end{equation}
and $C$ is an integration constant.  Since $C$ is an adiabatic constant,
the value does not change during the mass loss.  Thus we obtain the
density at the final state,
\begin{equation}
 \frac{y_{f}}{y_{i}}=\frac{C/z_{f}^{4}+q(z_{f})}{y_{i}}=
\frac{1}{y_{i}}
\left[(y_{i}-q(z_{i}))\frac{z_{i}^{4}}{z_{f}^{4}}+q(z_{f})\right],
\label{eqn:yfyi}
\end{equation}
and
\begin{equation}
 C=z^{4}[y-q(z)]=z_{i}^{4}[y_{i}-q(z_{i})].
\label{eqn:const}
\end{equation}
From equations (\ref{eqn:viriali}) and (\ref{eqn:virialf}), the velocity
dispersion at the final state is obtained as
\begin{equation}
 \frac{\sigma_{f}}{\sigma_{i}}=\left[
\frac{y_{f}z_{f}^{2}+bf(z_{f})/a}{y_{i}z_{i}^{2}+bf(z_{i})/a}
\right]^{1/2}.
\label{eqn:vel}
\end{equation}

In the next section, we derive the dependence of $y$ on $z$ by
specifying the function $f(z)$ in each case of density distributions of
baryon and dark matter.  Before studying such specific cases, we see the
limiting cases of no dark matter and negligible baryonic matter in the
next subsection.

\subsection{Limiting cases}\label{sec:limit}
First, we consider the case of self-gravitating system with no dark halo,
that is, $\rho_{d}/\rho_{i}\to 0$.  The final form is expected to be the
same as the result in \citet{ya87}.  For instantaneous gas removal, we
obtain from equation (\ref{eqn:diff2}),
\begin{equation}
 \frac{r_{f}}{r_{i}}\to\frac{M_{f}/M_{i}}{2M_{f}/M_{i}-1},
\end{equation}
where $M_{f}/M_{i}=\rho_{f}r_{f}^{3}/\rho_{i}r_{i}^{3}$.  This is the
well-known result that the system becomes unbound if a half of the mass
is removed.  On the other hand, for adiabatic gas removal, the second
term in equation (\ref{eqn:diff}) vanishes and then
\begin{equation}
 \frac{\rho_{f}}{\rho_{i}}\to\frac{r_{i}^{4}}{r_{f}^{4}}.
\end{equation}
The velocity dispersion is also obtained from equation (\ref{eqn:vel}),
\begin{equation}
 \frac{\sigma_{f}}{\sigma_{i}}\to\left[
\frac{\rho_{f}r_{f}^{2}}{\rho_{i}r_{i}^{2}}
\right]^{1/2}=\frac{r_{i}}{r_{f}}.
\end{equation}

Next, we consider the opposite case, $\rho_{i}/\rho_{d}\to 0$.  As
pointed out by \citet{ds86}, the size and velocity dispersion of
galaxies do not change during the mass loss because the gravitational
potential is dominated by dark matter.  Since the third term in
equation (\ref{eqn:diff2}) overwhelms other terms, $z$ does not
change against the change of $y$.  Thus the velocity dispersion also
does not change as obtained by equation (\ref{eqn:vel}).  This case
corresponds to that of \S III-c in \citet{ds86}.

Therefore, we confirm that our formulation is applicable to the limiting
cases of no dark matter and negligible baryonic matter.

\section{DYNAMICAL RESPONSE ON SIZE AND VELOCITY DISPERSION IN SPECIFIC CASES} 
In this section, we consider some specific density distributions.  As for the
baryonic component, we adopt the following two distributions.  One is the
Jaffe model \citep{jaffe} that approximates the de Vaucouleurs profile
when projected,
\begin{equation}
\rho(r)=\frac{4\rho_{b}r_{b}^{4}}{r^{2}(r+r_{b})^{2}},
\end{equation}
where $\rho_{b}$ and $r_{b}$ are the characteristic density and radius,
respectively.  Another is a model that approximates the exponential
profile when projected,
\begin{equation}
\rho(r)=\rho_{b}\sqrt{\frac{r_{b}}{r}}
\exp\left[-g_{\rm exp}\left(\frac{r}{r_{b}}-1\right)\right].
\end{equation}
The details of comparisons of these approximate profiles with the de
Vaucouleurs and exponential profiles are given in Appendix A.  As for
the dark matter halo, we consider the following two distributions for 
truncated isothermal and homogeneous spheres,
\begin{eqnarray}
\rho(r)&=&\rho_{d}\frac{r_{d}^{2}}{r^{2}}\theta(r_{d}-r),\qquad ({\rm isothermal})\\
\rho(r)&=&\rho_{d}\theta(r_{d}-r),\qquad ({\rm homogeneous})
\end{eqnarray}
respectively, where $\rho_{d}$ is a characteristic density of dark
matter and $\theta(x)$ is the Heaviside step function.  Finally we also
consider the NFW profile \citep{nfw},
\begin{equation}
\rho(r)=\rho_{d}c^{3}\left[\frac{cr}{r_{d}}\left(1+\frac{cr}{r_{d}}\right)^{2}\right]^{-1},
\end{equation}
where $c$ is the concentration parameter.  Although \citet{fm97, fm01}
and \citet{mgqsl} suggest a modified NFW profile with steeper slope in
the inner core, $\rho\propto r^{-1.5}$, we consider only the original
NFW profile.

In a hierarchical clustering scenario, the isothermal profile should be
applied to mergers and starbursts occurring near the centre of dark
halo.  This situation is realized when the mergers with satellite
galaxies are caused by the dynamical friction.  On the other hand, when
satellite-satellite mergers occur in the envelope of dark halo by random
collisions, the homogeneous profile should be applied.  Recent high
resolution $N$-body simulations suggest that many subhaloes survive and
are not disrupted \citep[e.g.,][]{oh99, m99, g00}.  If their internal
structures are also maintained, the isothermal profile should be applied
even in the case of mergers in the envelope of dark halo.  Moreover,
outer parts of individual subhaloes near the centre of their host halo
would be stripped by the tidal force from the host halo, then the mass 
ratio of baryon to dark matter would be high.  We also consider such a
case below.

\subsection{de Vaucouleurs-like distribution of baryon in an isothermal 
dark halo}

In this subsection, we investigate the dynamical response to mass
loss in the case of isothermal dark halo.  As already mentioned, we
adopt the Jaffe model which well reproduces the deprojected
three-dimensional de Vaucouleurs profile.  The corresponding mass
profile for baryon is
\begin{equation}
M_{b}(r)=\frac{16\pi\rho_{b}r_{b}^{3}r}{r+r_{b}},
\end{equation}
and for dark matter,
\begin{eqnarray}
M_{d}(r)&=&4\pi\rho_{d}r_{d}^{3}\left[\frac{r}{r_{d}}-\left(\frac{r}{r_{d}}-1
\right)\theta(r-r_{d})\right].
\end{eqnarray}
In the following, we use a parameter $m$ that stands for the mass ratio
of baryon to dark matter,
\begin{equation}
 m\equiv\frac{M_{b}(r\to\infty)}{M_{d}(r\to\infty)}.
\label{eqn:defm}
\end{equation}
From the self- and interaction-potential energies of baryon, we obtain 
the coefficients $a$ and $b$ in equation (\ref{eqn:genpot}) as follows,
\begin{equation}
a=128\pi^{2}G,\quad\mbox{and}\quad b=64\pi^{2}G,
\end{equation}
and the function $f(z)$ is
\begin{equation}
 f(z)=\frac{\ln(1+z)}{z}+\ln\left(1+\frac{1}{z}\right).
\end{equation}
Substituting these quantities into equation (\ref{eqn:q})
for adiabatic gas removal and into equation
(\ref{eqn:inst}) for instantaneous gas removal, we obtain the
dynamical response on size to gas removal.  Moreover, using equation
(\ref{eqn:vel}), we also obtain the response on velocity dispersion.
Because the corresponding equations and solutions are complicated, we
summarize those in Appendix \ref{sec:spec}.

We examine how the dynamical response depends on the ratio of initial
mass of baryon relative to dark matter halo $m_{i}\equiv M_{i}/M_{d}$ 
and the ratio of initial radius of baryonic component
relative to dark matter halo $z_{i}\equiv r_{i}/r_{d}$.  In the
following we consider two representative values of $m_{i}=0.2$ and 0.05,
and $z_{i}=0.2$ and 0.05, respectively.  In a standard cosmological
model, the baryon fraction is about 0.1, so $m_{i}\simeq 0.1$ is
expected if gas cooling is sufficiently effective.  Thus the two values
of $m_{i}$ correspond to two extreme cases in realistic situations.  
The size of disk galaxies is usually considered to be a radius at which 
the disks are supported by their rotation under an assumption that their
specific angular momenta conserve while the gas shrinks due to the cooling.  
The initial value of the specific angular momentum before the cooling is 
given by the so-called dimensionless spin parameter $\lambda$, which
distributes log-normally around 0.05, derived by the tidal torque due to
the large-scale density inhomogeneity in the Universe \citep{w84, ct96a,
ct96b, ng97}.  Because an effective radius of disk is $\simeq \lambda
r_{\rm vir}$ \citep{f83} and the size of elliptical galaxies is not
significantly different from the disk size, the values of $z_{i}=0.05$
and $0.2$ are realistic.  If outer parts of individual subhaloes are
tidally stripped by their host halo, even larger values might be
realized.

\begin{figure}
\epsfxsize=\hsize
\epsfbox{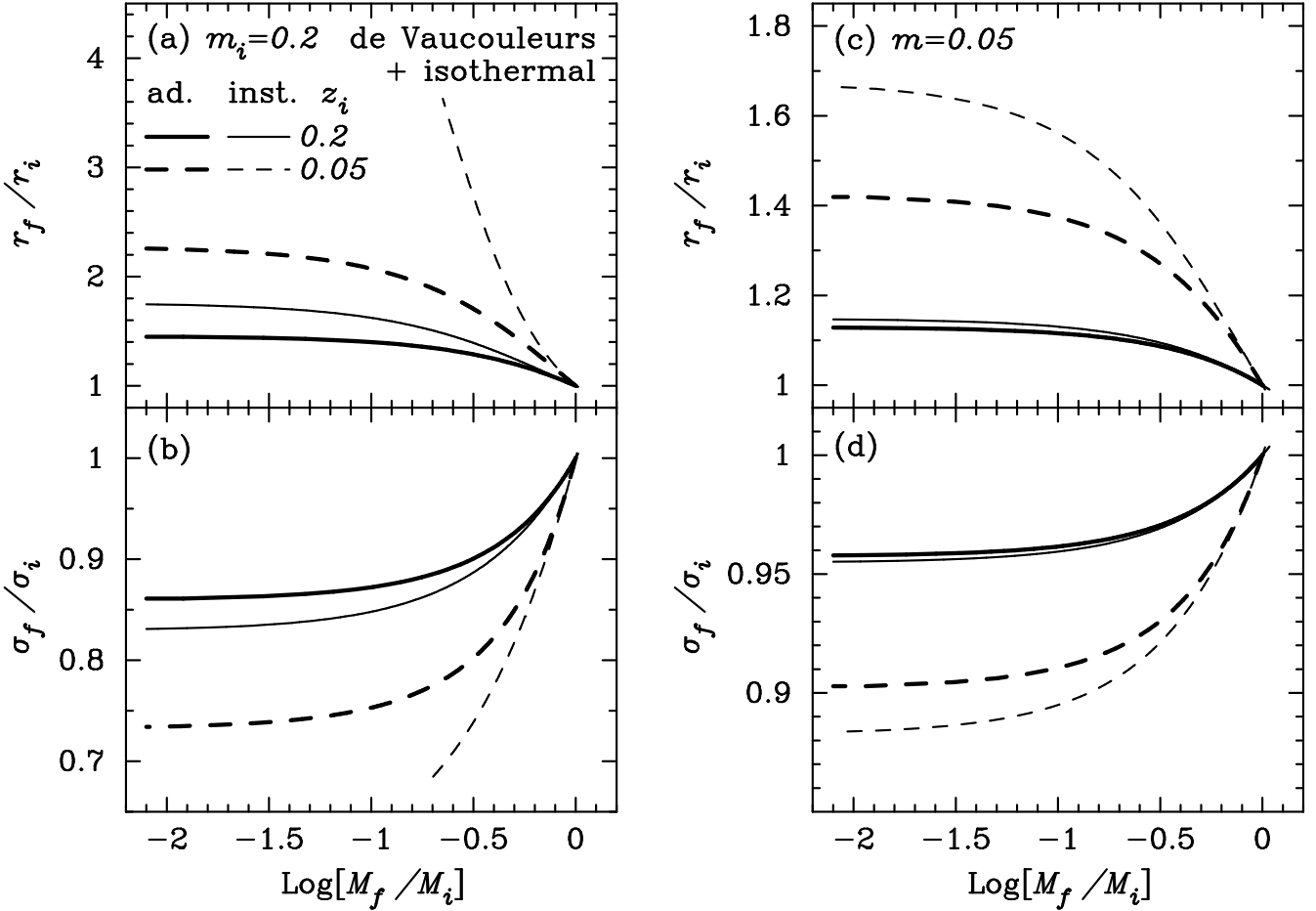}

\caption{Dynamical response of the baryonic component of deprojected de
 Vaucouleurs profile embedded in an isothermal halo, for various values
 of initial mass $m_{i}\equiv M_{i}/M_{d}$ and size $z_{i} \equiv
 r_{i}/r_{d}$.  The horizontal axis is the ratio of final to initial
 baryon masses.  Upper and lower panels indicate the changes of size and
 velocity dispersion, respectively, for $m_{i}=0.2$ (left panel) and
 0.05 (right panel).  The thick solid and thick dashed lines in each
 panel show the results of $z_{i}=0.2$ and 0.05, respectively, for
 adiabatic gas removal.  The thin solid and thin dashed lines show those
 results for instantaneous gas removal.  Note that the scales of
 vertical axes of left and right panels are different.}

\label{fig:dViso}
\end{figure}

In the left panels of Figure \ref{fig:dViso}, we plot the responses on
size (upper panel) and velocity dispersion (lower panel) for $m_{i}=0.2$.  
The thick solid and dashed lines indicate the results of $z_{i}=0.2$ and 
0.05, respectively, for adiabatic gas removal.  In the latter case of 
$z_{i}=0.05$, because the baryonic component is compact and gravitationally 
dominates in the central region, the changes of size and velocity dispersion 
are larger than in the former case of $z_{i}=0.2$.  In the right panels, 
we show the results of $m_{i}=0.05$.  The changes are smaller when compared 
to those for $m_{i}=0.2$ because of the small mass of baryonic component.  
Note that the changes for $(m_{i}, z_{i})=(0.2, 0.2)$ and $(0.05, 0.05)$ 
are very similar to each other.  The reason is, as pointed out by
\citet{nef}, that the changes can be measured by the value of $m_{i}/z_{i}$.  
If $m_{i}/z_{i}$ is sufficiently small as shown by the thick solid line in 
the right panels ($m_{i}/z_{i}=0.25$), the changes are negligible because 
the mass of dark matter dominates over the baryonic component.  In such a
case the assumption of dominant halo by \citet{ds86} can be justified.

For reference, we show the results for instantaneous gas removal by
thin lines.  As well known, the changes are greater than those for
adiabatic gas removal.  This tendency holds even in the dark
matter-dominated case.  Note that our results for this case are obtained
under the assumption of homologous change, which might not be justified
because some masses become unbound and leave the system.  Actually, by
using $N$-body simulations, \citet{nef} showed that some particles
initially in the central region of a dark halo become unbound when a
part of mass is instantaneously removed.

\subsection{Dependence on dark matter distribution}
In this subsection, we examine how the dynamical response depends on the 
density distribution of dark matter component, while the density distribution 
of baryonic component is fixed to the Jaffe model.

First we consider that the dark matter distributes homogeneously.  The
mass of dark matter within radius $r$ is
\begin{eqnarray}
M_{d}(r)&=&\frac{4\pi}{3}\rho_{d}r_{d}^{3}\left[\frac{r^{3}}{r_{d}^{3}}-\left
(\frac{r^{3}}{r_{d}^{3}}-1\right)\theta(r-r_{d})\right].
\end{eqnarray}
Then the coefficient $b$ is
\begin{equation}
b=\frac{64}{3}\pi^{2}G,
\end{equation}
and the function $f(z)$ is
\begin{equation}
 f(z)=\frac{1-2z}{2}+\frac{\ln(1+z)}{z}+z^{2}\ln\left(1+\frac{1}{z}\right).
\end{equation}
In the same manner as before, we obtain the dynamical response on
size and velocity dispersion for this case.

Next we consider the NFW profile.  The mass within radius $r$ is
\begin{equation}
 M_{d}(r)=4\pi\rho_{d}r_{d}^{3}\left[
 \ln\left(1+\frac{cr}{r_{d}}\right)
  -\frac{cr/r_{d}}{1+cr/r_{d}}\right].
\end{equation}
The coefficient $b$ is
\begin{equation}
 b=64\pi^{2}G,
\end{equation}
and the function $f(z)$ is
\begin{equation}
 f(z)=\frac{c\ln(cz)}{1-cz}+\frac{1}{z}\left[\ln(1-cz)\ln(cz)+{\rm Li}_{2}(cz)
\right],
\end{equation}
where Li$_{2}(z)$ is the dilogarithm.  In defining the parameter $m$ in
equation (\ref{eqn:defm}), we truncate $M_{d}(r)$ at $r=r_{d}$, because it
logarithmically diverges at $r\to\infty$.  Otherwise we do not assume
any truncation in the analysis.

\begin{figure}
\epsfxsize=\hsize
\epsfbox{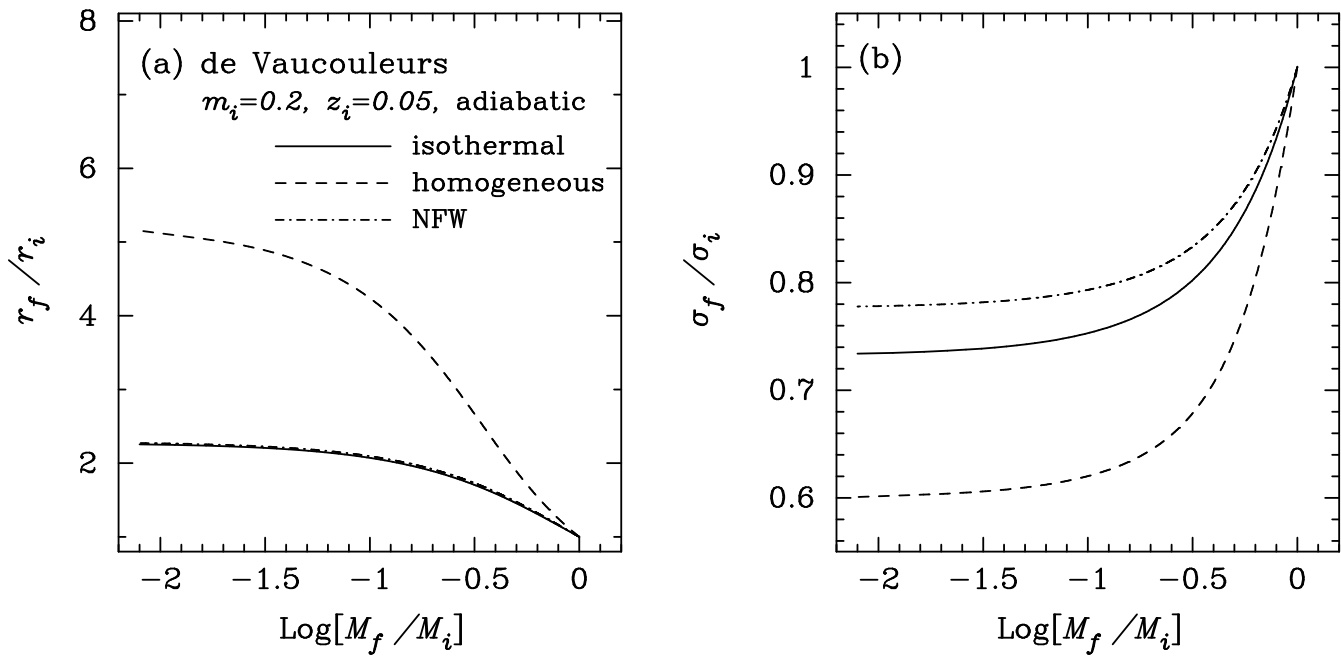}

\caption{Dynamical response of the deprojected de Vaucouleurs component
 within a dark halo with three different profiles.  The solid, dashed
 and dot-dashed lines denote the isothermal, homogeneous and NFW
 profiles, respectively.  Shown are the responses on size (left panel)
 and velocity dispersion (right panel) only for adiabatic gas removal 
 for the initial condition of $m_{i}=0.2$ and $z_{i}=0.05$.}

\label{fig:dVhom}
\end{figure}

The results for adiabatic gas removal are shown in Figure
\ref{fig:dVhom}.  The initial condition is $m_{i}=0.2$ and $z_{i}=0.05$
for all cases and $c=10$ for the NFW profile.  The solid, dashed and
dot-dashed lines indicate the isothermal, homogeneous and NFW profiles,
respectively.  In the case of the homogeneous dark halo, we find that
the changes of size and velocity dispersion are larger than those in the
case of the isothermal halo and the influence from dark halo is weaker.
This would reflects that the concentration of dark matter is lower than
that of the isothermal halo and that the gravitational potential energy
from dark matter is smaller, as indicated by the value of the
coefficient $b$.  This suggests that the size and velocity dispersion
can be significantly affected by mass loss even if surrounded by dark
matter, particularly when galaxy mergers frequently occur in the
envelope of dark halo.

In the case of the NFW halo with $c=10$, the responses are very similar 
to those for the isothermal halo.  This is because these haloes have 
much the same density distributions.

\subsection{Dependence on baryonic matter distribution}
In this subsection, we examine how the dynamical response depends on 
the density distribution of baryon within a given halo.  We consider 
the exponential-like distribution in the isothermal halo.  
The mass within $r$ is
\begin{equation}
 M_{b}(r)=\rho_{b}r_{b}^{3}\frac{\pi e^{g}}{g^{5/2}}\left[
-2\sqrt{\frac{gr}{r_{b}}}\left(2\frac{gr}{r_{b}}+3\right)\exp\left(-\frac{gr}
{r_{b}}\right)+3\sqrt{\pi}{\rm erf}\left(\sqrt{\frac{gr}{r_{b}}}\right)
\right],
\end{equation}
where we replace $g_{\rm exp}$ by $g$ for simplicity.  Then the
coefficients $a$ and $b$ are
\begin{equation}
 a=\frac{\pi^{2}(3\pi-4)e^{2g}}{g^{4}}G,\quad\mbox{and}\quad
 b=\frac{4\pi^{2}e^{g}}{g^{3}}G,
\end{equation}
and the function $f(z)$ is
\begin{equation}
 f(z)=-\frac{2g\exp(-g/z)}{\sqrt{z}}
+\sqrt{g\pi}\left\{
3\gamma+\frac{2g}{z}-\left(5+\frac{2g}{z}\right){\rm 
erf}\left(\sqrt{\frac{g}{z}}\right)+\ln64+3\ln\left(\frac{g}{z}\right)
\right\}+3\sqrt{z}G^{30}_{23}\left(\frac{g}{z}\left|^{3/2, 3/2}_{1/2, 1/2, 
1}\right.\right),
\end{equation}
where $\gamma$ is the Euler constant and $G_{pq}^{mn}$ is the Meijer
$G$ function (this symbol should not be confused with the gravitational 
constant $G$).  

\begin{figure}
\epsfxsize=\hsize
\epsfbox{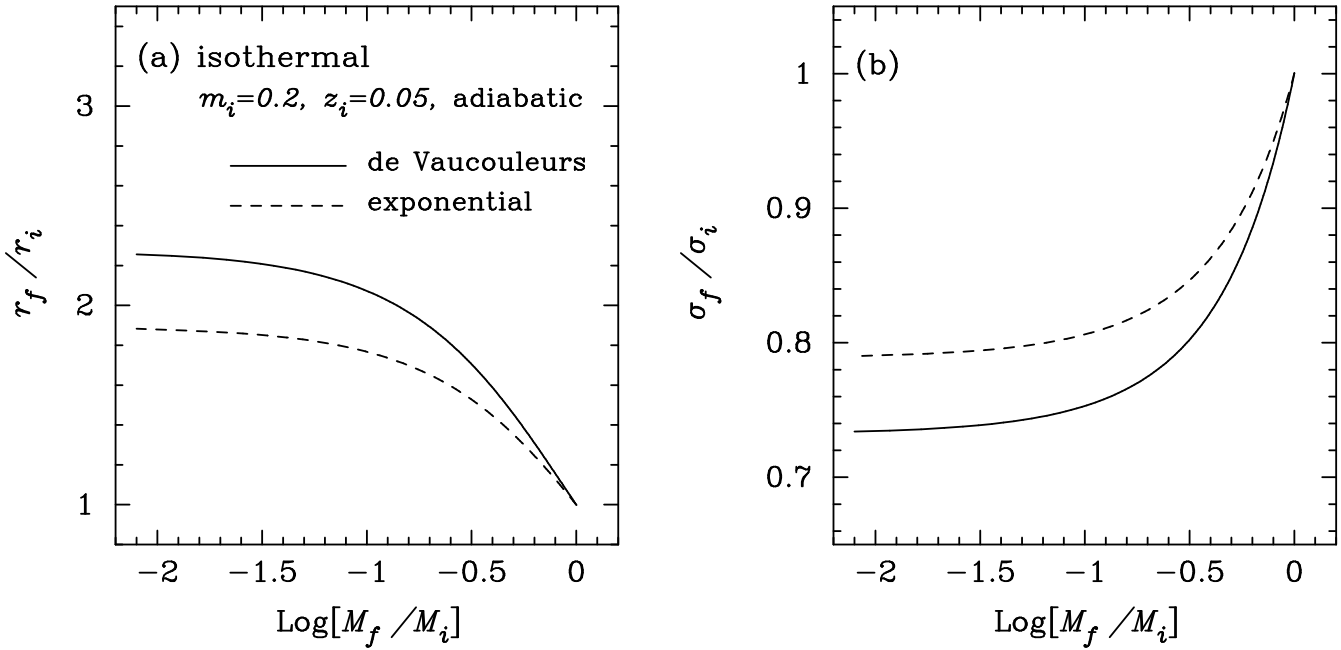}

\caption{Dynamical response of the baryonic component within the
isothermal dark halo.  The solid and dashed lines denote the de
Vaucouleurs and exponential profiles for the baryonic component,
respectively.  Shown are the responses on size (left panel) and 
velocity dispersion (right panel) only for adiabatic gas removal 
for the initial condition of $m_{i}=0.2$ and $z_{i}=0.05$.}

\label{fig:expiso}
\end{figure}

The results for adiabatic gas removal are shown in Figure
\ref{fig:expiso}.  Although we have calculated the exponential profile
of baryonic component in the homogeneous halo (Appendix
\ref{sec:exphom}), we do not show this case in the figure, because the
response is very similar to that for the de Vaucouleurs profile in 
the homogeneous halo.  The dependence on the density distribution of 
baryonic component is weaker compared to that of dark matter component, 
which suggests that the response mainly depends on the density distribution 
of dark matter.

\section{SUMMARY}

We analyzed the dynamical response on structural parameters of
elliptical galaxies to mass loss.  While in the previous analyses only 
two limiting cases of pure self-gravitating baryonic system and dark 
matter-dominated system were considered \citep[e.g.,][]{ya87, ds86}, 
we extended the formalism to the intermediate case between the two.  
Three-dimensional density distributions
of baryonic matter considered in this paper are the Jaffe model that
approximates the de Vaucouleurs profile and a model that approximates
the exponential profile in two dimensions.  As for the density distribution 
of dark matter, a truncated isothermal sphere and a homogeneous sphere are 
considered.  As a more realistic case, we also considered the NFW dark 
matter halo which surrounds the baryonic component having de Vaucouleurs 
profile.  Thus we found, from five combinations of baryon and dark matter,
that the response strongly depends on the central concentration of dark 
matter, reflecting the fraction of baryon relative to dark matter in the 
central region.  In contrast to the simple assumption by \citet{ds86} that 
structural parameters are not affected by mass loss, size and velocity 
dispersion can change even in dark matter, if baryon has sufficiently 
shrunk and become dense compared to dark matter at the onset of mass 
loss.

In this paper we considered both adiabatic and instantaneous gas
removal.  The adiabatic removal is justified if the timescale of gas
removal is sufficiently slow compared to the dynamical timescale of the
system.  On the other hand, in the case of instantaneous gas removal,
the effects of gas removal are stronger.  For example, when more than a
half of the mass is removed instantaneously, the pure self-gravitating
system becomes unbound.  Thus, our results for adiabatic removal place
a lower limit on the dynamical response to mass loss.  In order to check
the opposite case, we also derived the results for instantaneous gas
removal.  As expected, the changes of size and velocity dispersion are
larger than those for adiabatic gas removal.  Such drastic changes may
not validate our simple assumptions of homologous expansion of baryon
and static distribution of dark matter.

So far some scaling relations such as the velocity dispersion-magnitude 
relation have been discussed considering the dynamical response to mass 
loss in the context of monolithic cloud collapse scenario for formation 
of elliptical galaxies,  However, since the resuting structural change 
of baryonic component is found to depend significantly on the density 
distribution of dark matter, such scaling relations in the hierarchical 
clustering scenario should necessarily be reconsidered.  Our results 
suggest that the dispersed properties of dwarf/compact ellipticals 
would be well reproduced by a variety of their formation histories and 
environment, when taking into account mass-dependent gas removal.

In the hierarchical clustering scenario, galaxy formation processes are
rather complicated.  There are many physical processes such as radiative
gas cooling and galaxy merger as well as star formation and SN feedback.
It should be noted that the physical mechanism of galaxy merger would be
dynamical friction and/or random collision, while the mechanism of dark
halo merger is growth of density fluctuation.  In order to investigate
the observed properties of galaxies, we need to fully incorporate the 
effects of dynamical response into realistic models of galaxy formation
like our SAMs.  We will discuss various observables of ellipticals in 
more realistic situations in a forthcoming paper.

\section*{ACKNOWLEDGMENTS}    
This work has been supported in part by the Grant-in-Aid for the
Center-of-Excellence research (07CE2002) of the Ministry of Education,
Culture, Sports, Science and Technology of Japan.  Numerical computation
in this work was partly carried out at the Astronomical Data Analysis
Center of the National Astronomical Observatory and at the Yukawa
Institute Computer Facility.  The authors are grateful to the anonymous
referee for helpful comments to improve the text and the figures.

\appendix
\section{Simple models of density distribution}

\subsection{Approximate density distribution for the de Vaucouleurs $r^{1/4}$ 
profile}
Most of elliptical galaxies have the so-called
de Vaucouleurs $r^{1/4}$ luminosity profile,
\begin{equation}
\mu(R)=\mu_{e}\exp\left[-g_{\rm dV}\left\{\left(\frac{R}{R_{e}}
\right)^{1/4}-1\right\}\right],
\end{equation}
where $g_{\rm dV}=7.67$, $R$ is the distance from the centre of
luminosity distribution, and $R_{e}$ is the effective half-light radius,
at which the surface brightness reaches $\mu_{e}$.  In this paper,
assuming that mass traces luminosity, we regard the surface brightness
profile $\mu(R)$ as a surface mass density profile.

Since the de Vaucouleurs $r^{1/4}$ profile is two-dimensional projected
density distribution, we need to reconstruct three-dimensional density
distribution.  A direct procedure is to use the Abel inversion between a
surface density profile $\mu(R)$ and a three-dimensional density 
distribution $\rho(r)$,
\begin{equation}
\mu(R)=2\int_{R}^{\infty}\frac{\rho(r)r}{\sqrt{r^{2}-R^{2}}}dr,
\label{eqn:abel2}
\end{equation}
\begin{equation}
\rho(r)=-\frac{1}{\pi}\int_{r}^{\infty}\frac{d\mu(R)}{dR}
 \frac{dR}{\sqrt{R^{2}-r^{2}}}.
\label{eqn:abel}
\end{equation}
However, since the obtained profile is generally not given by an
analytic function, it is too complicated for analyses to use the exact
profile.  Therefore we adopt a simple, analytic model approximating the 
de Vaucouleurs profile when projected, that is, the Jaffe model
\citep{jaffe}.  By introducing characteristic density $\rho_{b}$ and
size $r_{b}$ of baryonic matter, the density distribution $\rho(r)$ is
written as follows,
\begin{equation}
\rho(r)=\frac{4\rho_{b}r_{b}^{4}}{r^{2}(r+r_{b})^{2}},
\end{equation}
where the half-mass radius $r_{e}$ in three-dimensional distribution
coincides with $r_{b}$.  The normalization of the density distribution
is given by equating the total masses of the de Vaucouleurs and the
Jaffe models.  From this condition, we find
\begin{equation}
\rho_{b}=\frac{2520\exp(g_{\rm dV})R_{e}^{2}}{g_{\rm dV}^{8}r_{b}^{3}}
\mu_{e}.  
\end{equation}
Moreover, we also find that the relationship between three- and
two-dimensional half-mass radii, $r_{e}$ and $R_{e}$, is $R_{e}\simeq
0.744r_{e}$.  These relations lead to $\mu_{e}\simeq 4\rho_{b} r_{e}$.

\begin{figure}
\epsfxsize=\hsize
\epsfbox{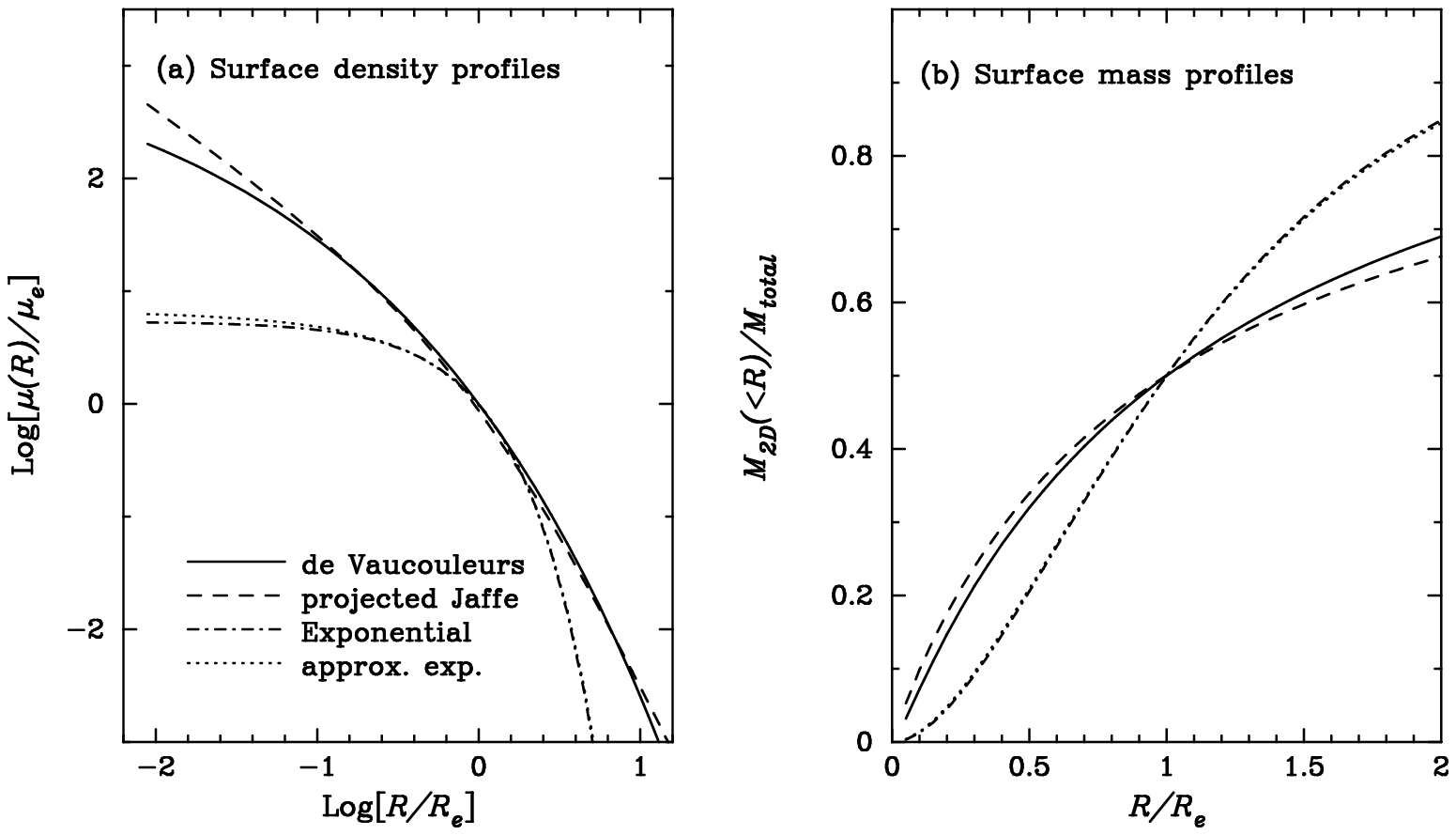}

\caption{(a) Surface density plotted against $R$.  Ths solid, dashed,
 dot-dashed, and dotted lines indicate the de Vaucouleurs profile,
 projected Jaffe model, exponential profile, and our approximated
 exponential model, respectively.  (b) Surface mass within $R$.  Lines
 are the same as in (a).  Note that our approximated exponential model
 agrees fairly well with the exponential profile.}

\label{fig:jaffe}
\end{figure}

The profile of surface density $\mu(R)$ and the profile of surface mass 
$M_{2D}(<R)$ within $R$ are shown in Figure \ref{fig:jaffe}.  The solid
and dashed lines indicate the de Vaucouleurs profile and projected Jaffe
model, respectively.  We adopt the normalizations of density and size
given by the above equations.  As proved by many authors, the Jaffe
model well reproduces the de Vaucouleurs profile.

\subsection{Approximate density distribution for the exponential profile}
Spiral galaxies and dwarf elliptical galaxies have an exponential
luminosity profile,
\begin{equation}
\mu(R)=\mu_{e}\exp\left[-g_{\rm exp}\left(\frac{R}{R_{e}}
-1\right)\right],
\end{equation}
where $g_{\rm exp}=1.68$.  Assuming the spherical symmetry, the
deprojected three-dimensional density distribution using equation
(\ref{eqn:abel}) is expressed by the modified Bessel function of the
second kind, $\mu_{e}g_{\rm exp}\exp(g_{\rm exp})K_{0}(g_{\rm
exp}r/R_{e})/\pi R_{e}$.  However, because of simplicity, we use an
approximate form for three-dimensional density distribution.  Note that 
the modified Bessel function of the second kind $K_{0}$ is asymptotically
proportional to $\exp(-g_{\rm exp}r/R_{e})/\sqrt{r}$ at $r\to\infty$.
Then, to obtain a simple form expressed in elementary functions, we
adopt the following density distribution
\begin{equation}
\rho(r)=\rho_{b}\sqrt{\frac{r_{b}}{r}}
\exp\left[-g_{\rm exp}\left(\frac{r}{r_{b}}-1\right)\right].\\
\end{equation}
As mentioned in the previous subsection, equating the total masses of
these distributions, we find
\begin{eqnarray}
\rho_{b}&=&\frac{2g_{\rm exp}^{1/2}R_{e}^{2}}{3\pi^{1/2}r_{b}^{3}}\mu_{e},
\end{eqnarray}
and $R_{e}\simeq 0.978r_{b}$.  The half-mass radius $r_{e}$ for
three-dimensional distribution is given by $r_{e}\simeq 1.295r_{b}$, then
$R_{e}\simeq 0.755 r_{e}$.  The projected density and mass profiles are
also shown in Figure \ref{fig:jaffe}.  The dot-dashed and dotted lines
indicate the exponential profile and our approximated exponential model,
respectively.  It is found that the exponential profile is well reproduced 
by our model.

\section{Derivations of Dynamical Response in Specific Cases}\label{sec:spec}
In this section we derive the relation between density and size for
the adiabatic change.  [The instantaneous change can directly be solved by
substituting the function $f(z)$ into equation (\ref{eqn:inst}).]

\subsection{de Vaucouleurs-like distribution of baryon in the isothermal 
dark halo}

Using the density and mass distributions of baryon and dark matter, 
the self- and interaction-potential energies of baryon in the isothermal 
halo of dark matter are calculated by
\begin{eqnarray}
-W_{\rm dV}^{\rm self}&=&128\pi^{2}G\rho_{b}^{2}r_{b}^{5},\\
-W_{\rm dV,iso}^{\rm int}&=&64\pi^{2}G\rho_{b}\rho_{d}r_{b}^{3}r_{d}^{2}\left[
\frac{\ln(1+z)}{z}+\ln\left(1+\frac{1}{z}\right)\right],
\end{eqnarray}
respectively.  From equation (\ref{eqn:diff}), the differential equation
for adiabatic gas removal is
\begin{equation}
 \frac{dy}{dz}=-4\frac{y}{z}-\frac{\ln(1+z)}{2z^{4}},
\end{equation}
then we obtain a solution
\begin{equation}
 y=\frac{C}{z^{4}}-\frac{1}{2z^{4}}\left[
-z+(1+z)\ln(1+z)\right],
\end{equation}
where $C$ is an integration constant given by equation
(\ref{eqn:const}).  In terms of mass instead of density, the response
can be expressed by
\begin{equation}
 m=4yz^{3},
\end{equation}
where $m$ denotes the mass ratio of baryon to dark matter.

\subsection{de Vaucouleurs-like distribution of baryon in the homogeneous dark halo}
The interaction-potential energies of the de Vaucouleurs-like profile of
baryon in the homogeneous halo of dark matter is
\begin{equation}
-W_{\rm dV,hom}^{\rm 
int}=\frac{64\pi^{2}}{3}G\rho_{b}\rho_{d}r_{b}^{3}r_{d}^{2}\left[
\frac{1-2z}{2}+\frac{\ln(1+z)}{z}+z^{2}\ln\left(1+\frac{1}{z}\right)\right].
\end{equation}
The differential equation for adiabatic gas removal is
\begin{equation}
 \frac{dy}{dz}=-4\frac{y}{z}-\frac{1}{6z^{2}}\left[
2-\frac{1}{z}+\frac{\ln(1+z)}{z^{2}}-2z\ln\left(1+\frac{1}{z}\right)
\right].
\end{equation}
Then we obtain
\begin{equation}
 y=\frac{C}{z^{4}}+\frac{1}{24z^{4}}\left[
6z+z^{2}-2z^{3}+2z^{4}\ln\left(1+\frac{1}{z}\right)-2(3+2z)\ln(1+z)
\right],
\end{equation}
where $C$ is an integration constant.  In terms of mass, we obtain
\begin{equation}
 m=12yz^{3}.
\end{equation}

\subsection{Exponential-like distribution of baryon in the isothermal dark halo}
The self-potential energy for baryon is given by
\begin{equation}
-W_{\rm exp}^{\rm 
self}=\frac{\pi^{2}(3\pi-4)e^{2g}}{g^{4}}G\rho_{b}^{2}r_{b}^{5},
\end{equation}
and the interaction-potential energy for baryon embedded in the dark halo is
\begin{eqnarray}
-W_{\rm exp,iso}^{\rm int}&=&
G\rho_{b}\rho_{d}r_{b}^{3}r_{d}^{2}\frac{4\pi^{2}e^{g}}{g^{3}}\nonumber\\
&&\times\left[
-\frac{2g\exp(-g/z)}{\sqrt{z}}
+\sqrt{g\pi}\left\{
3\gamma+\frac{2g}{z}-\left(5+\frac{2g}{z}\right){\rm 
erf}\left(\sqrt{\frac{g}{z}}\right)+\ln64+3\ln\left(\frac{g}{z}\right)
\right\}+3\sqrt{z}G^{30}_{23}\left(\frac{g}{z}\left|^{3/2, 3/2}_{1/2, 1/2, 
1}\right.\right)
\right].
\end{eqnarray}
From this, the differential equation becomes
\begin{equation}
  \frac{dy}{dz}=-4\frac{y}{z}+\frac{4g^{3}e^{-g}}{(3\pi-4)z^{9/2}}\left[
-2\sqrt{\frac{\pi z}{g}}+6\frac{z}{g}\exp\left(-\frac{g}{z}\right)+
\sqrt{\pi}\left\{2\sqrt{\frac{z}{g}}-3\left(\frac{z}{g}\right)^{3/2}\right\}
{\rm erf}\left(\sqrt{\frac{g}{z}}\right)
\right],
\end{equation}
then integrating the differential equation, we obtain
\begin{equation}
 y=\frac{C}{z^{4}}+\frac{4g^{7/2}e^{-g}}{(3\pi-4)z^{4}}\left[
\exp\left(-\frac{g}{z}\right)\left\{-2\sqrt{\frac{z}{g}}+3\left(\frac{z}{g}\right)^{3/2}\right\}+2\sqrt{\pi}\left\{
1-\frac{z}{g}-\left(1-\frac{z}{g}+\frac{3}{4}\frac{z^{2}}{g^{2}}\right)
{\rm erf}\left(\sqrt{\frac{g}{z}}\right)\right\}\right],
\end{equation}
where $C$ is an integration constant.  In terms of mass, we obtain
\begin{equation}
 m=\frac{3\sqrt{\pi}e^{g}}{4g^{5/2}}yz^{3}.
\end{equation}

\subsection{Exponential-like distribution of baryon in the homogeneous dark halo}\label{sec:exphom}
The interaction-potential energy for baryon embedded in the dark halo is
\begin{eqnarray}
-W_{\rm exp,hom}^{\rm int}&=&G\rho_{b}\rho_{d}r_{b}^{3}r_{d}^{2}\frac{\pi^{2}e^{g}}{6g^{5/2}}\nonumber\\
&&\times\left[
\frac{16g\sqrt{\pi}}{z}-2\exp\left(-\frac{g}{z}\right)\left\{
8\sqrt{\frac{g}{z}}-70\sqrt{\frac{z}{g}}-105\left(\frac{z}{g}\right)^{3/2}
\right\}-\sqrt{\pi}\left(105\frac{z^{2}}{g^{2}}-36+16\frac{g}{z}\right)
{\rm erf}\left(\sqrt{\frac{g}{z}}\right)
\right].
\end{eqnarray}
Therefore the coefficient $b$ is
\begin{equation}
 b=\frac{\pi^{2}e^{g}}{6g^{5/2}}G,
\end{equation}
and the function $f(z)$ is
\begin{equation}
 f(z)=\frac{16g\sqrt{\pi}}{z}-2\exp\left(-\frac{g}{z}\right)\left\{
8\sqrt{\frac{g}{z}}-70\sqrt{\frac{z}{g}}-105\left(\frac{z}{g}\right)^{3/2}
\right\}-\sqrt{\pi}\left(105\frac{z^{2}}{g^{2}}-36+16\frac{g}{z}\right)
{\rm erf}\left(\sqrt{\frac{g}{z}}\right).
\end{equation}
The differential equation for $y$ is
\begin{equation}
  \frac{dy}{dz}=-4\frac{y}{z}+\frac{g^{5/2}e^{-g}}{3(3\pi-4)z^{4}}\left[
\exp\left(-\frac{z}{g}\right)\left\{56+140\frac{z}{g}+210\left(\frac{z}{g}\right)^{2}\right\}-8\sqrt{\pi}+\sqrt{\pi}\left\{8-105
\left(\frac{z}{g}\right)
^{3}\right\}{\rm erf}\left(\sqrt{\frac{g}{z}}\right)
\right],
\end{equation}
then integrating the differential equation, we obtain
\begin{eqnarray}
 y&=&\frac{C}{z^{4}}+\frac{g^{7/2}e^{-g}}{6(3\pi-4)z^{4}}\nonumber\\
&&\times\left[
\exp\left(-\frac{g}{z}\right)\sqrt{\frac{z}{g}}\left(
-24+28\frac{z}{g}+70\frac{z^{2}}{g^{2}}+105\frac{z^{3}}{g^{3}}
\right)+8\sqrt{\pi}\left(3-2\frac{z}{g}\right)-
\frac{\sqrt{\pi}}{2}{\rm erf}\left(\sqrt{\frac{g}{z}}\right)\left(
48-32\frac{z}{g}+105\frac{z^{4}}{g^{4}}
\right)
\right],
\end{eqnarray}
where $C$ is an integration constant.  In terms of mass, we obtain
\begin{eqnarray}
 m=\frac{9\sqrt{\pi}e^{g}}{4g^{5/2}}yz^{3}.
\end{eqnarray}

\subsection{de Vaucouleurs-like distribution of baryon in the NFW dark halo}
The NFW density distribution is written as 
\begin{equation}
\rho(r)=\rho_{d}c^{3}\left[\frac{cr}{r_{d}}\left(1+\frac{cr}{r_{d}}\right)^{2}\right]^{-1},
\end{equation}
where $c$ is the concentration parameter.  The mass within radius $r$ is
\begin{equation}
 M_{d}(r)=4\pi\rho_{d}r_{d}^{3}\left[\ln\left(1+\frac{cr}{r_{d}}\right)
  -\frac{cr/r_{d}}{1+cr/r_{d}}\right].
\end{equation}
The interaction-potential energy is
\begin{eqnarray}
-W_{\rm dV,NFW}^{\rm int}&=&
64\pi G\rho_{b}\rho_{d}r_{b}^{3}r_{d}^{2}
\left[\frac{c\ln(z)}{1-cz}+\frac{\ln(1-cz)\ln(cz)+{\rm Li}_{2}(cz)}{z}\right],
\end{eqnarray}
where Li$_{2}(z)$ is the dilogarithm, $\int_{z}^{0}[\ln(1-t)/t]dt$.
The differential equation is
\begin{equation}
 \frac{dy}{dz}=-4\frac{y}{z}+\frac{1}{2z^{3}}\left[
\frac{c}{1-cz}-\frac{c(1-2cz)}{(1-cz)^{2}}\ln(cz)-\frac{\ln(1-cz)\ln(cz)}{z}-\frac{{\rm Li}_{2}(cz)}{z}\right],
\end{equation}
then integrating this equation, we obtain
\begin{eqnarray}
y&=&\frac{C}{z^{4}}+\frac{1}{2cz^{4}}\left[-4cz+(4-cz)\ln(1-cz)\ln(cz)+
\frac{cz(4-3cz)\ln(cz)}{1-cz}+(4-cz){\rm Li}_{2}(cz)\right],
\end{eqnarray}
where $C$ is an integration constant.  In terms of mass, we obtain
\begin{eqnarray}
m=\frac{4yz^{3}}{\ln(1+c)-c/(1+c)}.
\end{eqnarray}

\bsp
\end{document}